\documentclass[a4paper]{article}
\usepackage{a4wide}
\usepackage{latexsym}
\usepackage{graphicx,subfigure}
    \graphicspath{{./}{figure/}}
\usepackage{algorithm}
\usepackage{algpseudocode}
\usepackage{epstopdf}
    \graphicspath{{./}{figures/}}
\usepackage[colorlinks=true,linkcolor=blue,citecolor=blue]{hyperref}
\usepackage{xcolor}
\usepackage{amsfonts,amsmath,amsthm,mathtools,bbm,cool}
\usepackage{bm}

\newtheorem{theorem}{Theorem}[section]

\theoremstyle{remark}\newtheorem{remark}[theorem]{Remark}

\newcommand{\be}{\begin{equation}}
\newcommand{\ee}{\end{equation}}

\newcommand{\e}{\epsilon}

\newcommand{\fer}[1]{(\ref{#1})}
\newcommand{\R}{\mathbb R}
\newcommand{\N}{\mathbb N}
\allowdisplaybreaks
\def\be#1\ee{\begin{equation}#1\end{equation}}
\newenvironment{equations}{\equation\aligned}{\endaligned\endequation}
\begin{document}
\title{A kinetic description of the body size distributions of species}

\author{Stefano Gualandi \thanks{Department of Mathematics ``F. Casorati'', University of Pavia, Italy. {\tt stefano.gualandi@unipv.it}} \and
        Giuseppe Toscani \thanks{Department of Mathematics ``F. Casorati'', University of Pavia, and The Institute for Applied Mathematics and Information Technologies of CNR, Pavia, Italy. {\tt giuseppe.toscani@unipv.it}} \and 
      Eleonora Vercesi \thanks{Department of Mathematics ``F. Casorati'', University of Pavia, Italy. {\tt eleonora.vercesi01@universitadipavia.it }}
             }  
\date{}

\maketitle

\begin{abstract}
In this paper, by resorting to classical methods of statistical mechanics, we build a kinetic model able to reproduce  the observed statistical weight distribution of many diverse species. The kinetic description of the time variations of the weight distribution is based on elementary interactions that describe in a qualitative and quantitative way successive evolutionary updates, and determine explicit equilibrium distributions. Numerical fittings on mammalian eutherians of the order Chiroptera population illustrates the effectiveness of the approach. 

\medskip


\end{abstract}

\tableofcontents

%%%%%%%%%%%%%%%%%%%%%%%%%%%%%%%%%%%%%%%%%%%%%%%%%%%%%%%%%%
\section{Introduction}

In this paper, by means of classical methods of statistical mechanics, we present a qualitative and quantitative description of the possible mechanisms leading to the observed statistical weight distribution of species. 
There are several reasons for this choice, the main ones being that statistical mechanics provides a powerful approach to the mathematical modeling of systems composed of a huge number of agents interacting with each other and/or the environment, and has as its primary product the understanding of the relationship between parameters in microscopic rules and the resulting macroscopic statistical outcomes. In reason of this possibility, in the last two decades there has been a trend toward applications of statistical mechanics to interdisciplinary fields ranging from the classical biological context to new aspects of socio-economic phenomena \cite{ABG,BKS,BCKS,CFL,NPT,PT2}.

In order to justify at best the statistical mechanics approach, classically used for gas dynamics where the number of molecules in a mole is quantified by Avogadro's constant, the system under consideration must possess some minimum requirements. On the one hand, the population of growing organisms  it must be at least of the order of millions, and on the other hand, the time in which we observe the system must be  large enough to ensure that statistical equilibrium has been reached. Regarding the first requirement, among other possible choices, we will test our modeling outcomes with the population of Chiroptera. Indeed, the population of eutherian mammals of the order Chiroptera, which includes about 1100 species grouped in 18 families, is sufficiently numerous, uniformly distributed on planet earth, and intensively studied by naturalists \cite{Mammalogy}. For the second requirement, considering that the the estimated appearance of the species dates back to about fifty million years ago, and the generations have succeeded on average at the rate of few years, the statistical distribution observed to date is clearly close to equilibrium.

Let us briefly introduce the content of the paper. Following Ref. \cite{PT2}, let us consider a multi-agent system  in which growing organisms are characterized by their weight $w$, measured in some unit, and let $f(w, t)$ denote the statistical distribution of the population's weight at a certain time $t \ge 0$, normalized to have unit mass. The goal is to characterize the time evolution of the probability density $f(w,t)$ subject to changes obeying to objective mechanical and behavioral principles when interacting with the surrounding environment. 

To achieve this, let us fix a certain time $t= t_0$, considered as the initial time of the observations,  and let $ w_0 $ be the weight of a representative grower of the population at time $ t_0 $. Then, we evaluate the statistical distribution of the weight of the population at regular time intervals $ t_{n+1} $, with $n$ nonnegative integer, $ n\in \N_+$, consequent to    the elementary variation $w_n \to w_{n+1}$ of the weight.   This change is modeled as follows
\begin{equation}\label{coll1}
w_{n+1} = w_n^*= w_n  - \Phi^\epsilon \left(\frac{w_n}{w_T}\right) w_n +  \Psi^\e(w_n)\,w_n \eta.
\end{equation}
Thus, at each time step, the weight $w_n$  can be modified by two different mechanisms, expressed in mathematical terms by two multiplicative terms, both parameterized by a small positive parameter $\e \ll 1$:
\begin{itemize}
\item[$i)$] The transition function $\Phi^\epsilon (\cdot)$ characterizes the deterministic variations of the weight as a function of the quotient $w_n /w_T$, where $w_T$ is the target value of the weight to be achieved. This function is an asymmetric function, negative below the target value $w_T$ and positive above, that it is designed to take into account both the principles of ontogenetic growth \cite{West}, and the behavioral aspects linked to the growth itself \cite{DT2}.  The form of the $\Phi^\e$ function  reflects the fact that organisms change their weight with the goal of approaching the target value, and that this change requires asymmetric effort, depending on whether the current state is above or below the target. This choice is in agreement with the fact that, from bacteria to insects to mammals, a seemingly universal feature of sufficiently large taxonomic groups is a frequency distribution of body sizes among species that is highly right-skewed, even on a logarithmic scale \cite{Bon,May,Stan}.

\item[$ii)$] Random fluctuations due to unknown factors found in the environment are  expressed by the product between the \emph{linearly weight depending} random variable $w\eta$ and the function $\Psi^\e(\cdot)$.  The  choice is to consider that the random variable $\eta$ is of zero mean and bounded variance, given by $\langle \eta \rangle =0$, $\langle \eta_\e^2 \rangle  = \sigma$. The  function $\Psi^\e(\cdot)$ quantifies the amount of weight that can be achieved or lost due to the presence of the environment, in terms of the weight of the individual. The simplest choice is to assume that  $\Psi^\e(w)$, $w \ge 0$ is  a positive constant, namely that the random variation of weight are proportional to the weight itself. However this choice leads to the possibility to have infinite weight \cite{DT,DT2,GT1,GT2}. In this case we assume that the weight-related energy constraints are such that random fluctuations will be negligible above a certain weight level, thus imposing that the function  $\Psi^\e$ is a non-increasing function of bounded support, and consequently the weight can not exceed a certain value.
\end{itemize}

At difference with the other parameters characterizing the shape of the functions appearing in the elementary update \fer{coll1}, the meaning of the small parameter $\e\ll1$ is  directly linked to the size of the possible variations of weight  at each generational change. Indeed, it is realistic to assume that at each new generation  the changes in weight of the organisms under study is extremely small, and that only after an extremely large number of changes can the statistical weight distribution stabilize. 

It is important to remark that the idea of looking at the long-time effect of elementary variations (in our case described by the relation  \fer{coll1}) is strongly reminiscent  of the work of Sir Francis Galton \cite{Gal1,Gal2}, in particular of his bean machine,  a device invented to demonstrate the central limit theorem, in particular that with  a large enough sample the binomial distribution approximates a normal distribution. Indeed, as we shall see in some detail in Section \ref{sec:Galton}, Galton's experiment can be fruitfully described by resorting to classical kinetic methods, that allow to understand how to use the same arguments to obtain explicit results in the case of more complicated ways of interaction. 

Leaving details about the kinetic description of a repeated experiment to Section \ref{sec:Galton}, and postponing the characterization of the explicit form of the functions $\Phi^\e$ and $\Psi^\e$ to Section \ref{kinetic},  we recall that, starting from the definition of the elementary transition process  \fer{coll1}, the study of the
continuous in time-evolution of the statistical distribution $f(w,t)$ of the weight follows by
resorting to kinetic models \cite{Cer,PT2}.
 For any given value of the small parameter $\e$, the variation of the  density $f(w,t)$   obeys to a linear
Boltzmann-like equation, fruitfully written
in weak form. The weak form corresponds to say that the solution $f(w,t)$
satisfies, for all smooth functions $\varphi(w)$ (the observable quantities), the integro-differential equation
\begin{equation}\label{B1}
\dfrac{d}{dt} \int_{\mathbb R_+} \varphi(w) f(w,t)dw = \frac 1{\tau(\e)} \left \langle \int_{\mathbb R_+} \chi\left(\frac w{[m]}\right)\, (\varphi(w^*)-\varphi(w))f(w,t)dw \right\rangle.
\end{equation}
In \fer{B1}  we denoted  with $\langle \cdot \rangle$ the expectation with respect to the random parameter $\eta$ present in \fer{coll1}. The constant quantity $[m]$  is the unit measure we use to quantify the weight, while the positive function $\chi(\cdot)$ is a kernel characterizing the  frequency of the elementary growth transitions relative to the weight $w$. Since we are mainly interested on the stationary solutions to equation \fer{B1},  we will apply the so-called Maxwellian simplification \cite{Cer}, which consists of assuming that the kernel $\chi$ is a fixed positive constant, that for simplicity will be fixed equal to one. Indeed, as discussed in Ref. \cite{FPTT1}, the shape of the steady state solution does not depend on the details of the kernel function $\chi(\cdot)$. Last, $1/\tau(\e)$ is the time scale at which we can observe finite variations of the observable quantities in consequence of elementary interactions.

The integral on the right-hand side of equation \fer{B1} represents the balance in density  between growing organisms  that  modify their weight from $w$ to $w^*$ (loss term with negative sign), and, respectively, growing organisms  that  change their weight from from $w^*$ to  $w$  (gain term with positive sign).

Choosing $\varphi(w) = 1$ we immediately obtain that there  the integral of the statistical distribution $f$ is preserved in time, that is
\[
\int_{\mathbb R_+}  f(w,t)\, dw = \int_{\mathbb R_+}  f(w,t_0)\, dw,
\]
for any $t >0$. Consequently, the solution to \fer{B1} remains a probability density for all times, if it is so initially.

Once the continuous in time evolution of the density has been characterized, the main object to determine is the steady state distribution of the multi-agent system, which is obtained by solving equation \fer{B1} in which the time derivative is set equal to zero. While the explicit form of the steady states of the linear kinetic model  \fer{B1} is  available only in a very limited number of cases \cite{PT2}, explicit steady states can be obtained in the  case in which $\e \ll1$, which corresponds to assume that at any time step  the variations in weight of the population are very small, while, to observe a non negligible size of their effect,  the time scale  is modified according to the choice  $\tau(\e) = \e$. 

In this particular regime, known in classical kinetic theory with the name of \emph{grazing} regime \cite{FPTT,Vil},  the steady state solution to the kinetic equation \fer{B1} is close to the steady profile of the Fokker--Planck equation \cite{Vil}
 \be\label{FP1}
  \frac{\partial f(w,t)}{\partial t}= \frac {\sigma} 2 \frac{\partial^2 }{\partial w^2}
 \left(\Psi^{2}(w)\, w^2 f(w,t)\right ) + \frac{\partial}{\partial w}\left[w \, \Phi \left(\frac{w}{w_T}\right)f(w,t) \right],
 \ee
where the functions $\Phi$ and $\Psi$ in \fer{FP1} are defined by
\be\label{graz}
\Phi(w) = \lim_{\e \to 0} \frac 1\e \, \Phi^e (w), \quad \Psi(w) = \lim_{\e \to 0} \frac 1{\sqrt\e}\, \Psi^e (w).
\ee
It is immediate to recognize that both the function $\Phi$ characterizing the first order drift term and the function $\Psi$ quantifying the second order diffusion term of equation \fer{FP1} are closely related to details of the elementary interaction \fer{coll1}. Indeed
\be\label{dri}
\lim_{\e \to 0} \frac 1\e \, \Phi^e (w)\, w = \lim_{\e \to 0} \frac 1\e\left\langle w^* - w\right\rangle,
\ee
while
\be\label{diff}
\lim_{\e \to 0} \frac 1{\sqrt\e} \, \Phi^e (w) \, w = \lim_{\e \to 0} \frac 1{\sqrt\e} \,\sqrt{ \left\langle (w^* - w)^2 \right\rangle},
\ee
Hence, the drift term in \fer{FP1} is linked by the mean variation of the weight, and the diffusion coefficient in \fer{FP1} is  linked to the variance of the random fluctuations. It is important to remark that, unlikely, in the \emph{grazing} limit procedure various details of the elementary variation \fer{coll1} are lost. In particular, the random variable $\eta$ that characterizes  the unpredictable fluctuations of the weight of the organisms due to the environment appears in the Fokker--Planck equation only through its variance $\sigma$.

In reason of the choice of the function $\Psi^\e$, the explicit stationary solution of the Fokker--Planck equation is of bounded support, and in relevant cases it is given by a generalized Beta probability density of the first kind \cite{MX}, which, in a certain range of the parameters of the elementary interaction \fer{coll1}, perfectly fits the available data of the weight distribution of Chiroptera population. However, we outline once again that the methodology used in this paper can be easily extended to explain the statistical distribution of the body weight of other species of organisms.
Indeed, the approach proposed here is based on a Boltzmann-type model where the elementary variations of the weight  are determined by a general transition function which takes ontogenetic growth and random fluctuations into account. Further, by suitable choosing the parameters, the microscopic variations are coherent with most of the known growth models.  Related kinetic equations based on elementary variations of type \fer{coll1}, and inspired by early socio-economic motivations \cite{KT},  have been introduced recently  to justify the formation of lognormal distribution in collective phenomena in Refs. \cite{DT,DT2,GT1,GT2}.

Partial differential equations describing the biological evolution of species body masses within large groups of related species, e.g., terrestrial mammals, have been previously considered in the pertinent literature.  A related quantitative model has been recently derived in Ref. \cite{CR} (cf. also Refs. \cite{CH,CSR}). In this interesting approach  body mass evolves according to branching (speciation), multiplicative diffusion, and an extinction probability that increases logarithmically with mass. This evolution is described in terms of a convection-diffusion-reaction equation for the body mass, where, however, at difference with the present approach, the diffusion term is assumed to be constant.

In more detail, the paper is organized as follows. Section \ref{sec:Galton} is devoted to a brief introduction to Galton bean machine and to its connection with diffusion and linear kinetic equations. Then, Section \ref{kinetic} introduces the Boltzmann-type linear kinetic model for body mass growth, where elementary variations in body mass are considered that depend on a transition function that finds out deterministic, both mechanical and behavioral, variations in weight, and on random fluctuations due to the environment.  Then, the Fokker--Planck type equation \fer{FP1} is shortly derived in Section  \ref{sec:grazing}, where also its explicit equilibrium density is obtained. Last, numerical fittings on Chiroptera populations illustrating the effectiveness of the approach are presented in Section \ref{fitting}.

\section{The legacy of Galton bean machine} \label{sec:Galton}
 
 \subsection{The normal distribution}
Linear kinetic models can be easily described by resorting to Galton experiment \cite{Gal1,Gal2}. Indeed, Galton bean machine represents a clear visual approach to link  {repeated elementary interactions} to a consequent {universal} steady state density. In Galton board, balls subject to gravity exhibit repeated variations of direction before falling on the ground. Each symmetric variation is determined by interaction of the ball with the vertex of an isosceles triangle.  At the bottom, if {the number of balls is sufficiently high}, one recognizes the profile of the  {Gaussian} density 
 \[
 M(x) = \frac 1{\sqrt{2\pi\sigma}}\exp\left\{-\frac{x^2}{2\sigma} \right\}.
 \]
Galton experiment is equivalent to a sequence of \emph{Bernoulli trials} (or binomial trials), in which the trial is a random experiment with exactly two possible outcomes, {success} and {failure}. 
  The probability of success ({$1/2$} in Galton experiment) is the same every time the experiment is conducted. 
 The classical way to give a mathematical interpretation of the result is to pass from the discrete to the continuous description. In the half-plane $(x,t)$, $t\ge 0$  the {continuous} description of the time variation of the probability  $p(x,t)dx$  to find the ball in the interval $(x, x+\Delta x)$ at time $t +\Delta t \ge 0$ is expressed by
  \[
  p(x,t+\Delta t) = \frac 12 p(x+ \Delta x, t ) + \frac 12 p(x- \Delta x, t ).
  \]
Expanding in Taylor's series, at the first order in time the left-hand side and at the second order in position the right-hand side, shows that in the limit {$(\Delta x)^2/\Delta t \to \sigma>0$}, $p(x,t)$ satisfies the diffusion equation
 \[
 \frac{\partial p}{\partial t} = {\frac\sigma{2} }\frac{\partial^2 p}{\partial x^2}.
 \]
At this point, it is enough to observe that in Galton experiment the repeated trials start from $x=0$.  Consequently, if the number of repeated trials is sufficiently large, the resulting profile is well approximated by the \emph{source-type solution} of the diffusion equation.
   The {source-type solution} to the diffusion equation at time $t >0$ is the {Gaussian density}
 \[
 M(x, t) = \frac 1{\sqrt{2\pi\sigma t}}\exp\left\{-\frac{x^2}{2\sigma t} \right\}.
 \]
This gives an answer to the visual result of the experiment.

Galton experiment can be modeled in a slightly different way, by resorting to the classical approach of collisional kinetic theory. Let us consider a huge number of identical Galton bean machines, and let us run a trial of the experiment simultaneously on all machines.  Let us denote by $f(x,t)\, dx$, $x \in \R$,  the percentage of balls which at time $t\ge 0$  lie in the  interval $(x, x +dx)$. The quantity $f(x,t)$ changes in time since balls are subject to interactions which modify their position.  The {elementary interaction} is given by
  \be\label{ga-in}
  x_* = x + \sqrt\sigma \eta,
  \ee
 where $\eta$ is a {symmetric random variable} taking values $\pm 1$ with probability $1/2$ ( $2\sqrt\sigma$  is the length of the base of  the triangles). 
According to kinetic theory \cite{Cer,PT2}, the  evolution in time of the density $f(x,t)$  is obtained by looking at its time variation  along the \emph{observable quantities} $\varphi(x)$,  written as 
\be\label{bo}
\frac d{dt} \int_\R \varphi(x)f(x,t)\, dx = \frac 1\tau \left\langle \int_\R  \left[ \varphi(x_*) -\varphi(x)\right] f(x,t) \,dx\right\rangle.
 \ee
The integral on the right-hand side of equation \fer{bo} measures the variation of the observable quantity $\varphi$ consequent to the variation of the positions of the balls due to the elementary law \fer{ga-in}. The presence of the mean $\langle\cdot\rangle$ is due to the presence of the random variable $\eta$. The positive part of the integral quantifies the variation of $\varphi$ consequent to balls which change their position from $x_*$ to $x$ (gain term),  while the negative part of the integral quantifies the variation of $\varphi$ consequent to balls which change their position from $x$ to $x_*$ (loss term). Last, the time frequency $1/ \tau$ is a measure of the velocity at which the phenomenon relaxes towards the steady or self-similar solution \cite{Cer}.   The trial starts with balls located in $x=0$. The kinetic equation \fer{bo} allows to compute the time evolution of the main observable quantities, like the principal moments.

If we choose $\varphi(x) = 1$, we obtain
  \[ 
  \frac d{dt} \int_\R   \,f(x,t)\, dx =  0.
  \]
 Hence, the mass is preserved in time. This means that $f(x,t)$ remains a probability density for all $t \ge 0$, if it is so initially.
  Same conservation result if $\varphi(x) = x$, since {$\langle x_* -x \rangle = \langle \eta \rangle = 0$}.
  The mean value of $f(x,t)$, equal to zero at time $t=0$, is {equal to zero for all times $t \ge 0$}.
 
  However, since {$\langle x_*^2- x^2\rangle = \langle 2x\sqrt\sigma\eta + \sigma\eta^2\rangle = \sigma$},  
\[
\frac d{dt} \int_\R  x^2 f(x,t)\, dx =  \frac\sigma\tau  \int_\R  f(x,t)\, dx = \frac\sigma\tau.
 \]
 Consequently, the variance of the density function {diverges with time}. This shows that we can not expect that the solution will converge to a  steady state of finite variance. 
 
 This obstacle can be circumvented by suitably {shrinking the domain}. Given a positive constant $\lambda <1$ let us consider the modified interaction 
  \be\label{ga-sc}
 x_* = x(1-\lambda) + \sqrt\sigma\eta.
  \ee
The shrinking does not modify the characteristics of the shape of $f(x,t)$. Also, the mean value of $f(x,t)$  remains {equal to zero for all times $t \ge 0$}.
 However, now {$\langle x_* ^2 -x^2\rangle = \sigma - \lambda(2-\lambda)x^2$}. Consequently
 \be\label{vava}
\frac d{dt} \int_\R  x^2 f(x,t)\, dx =  \frac\sigma\tau \int_\R   f(x,t)\, dx  - \frac{\lambda(2-\lambda) }\tau \int_\R  x^2 f(x,t)\, dx.
 \ee
The variance of the density function {remains uniformly bounded in time}. Indeed, solving the differential equation we obtain
  \[
  \int_\R  x^2 f(x,t)\, dx =  e^{-\lambda(2-\lambda)/\tau} \,\int_\R  x^2 f_0(x)\, dx  + {\frac{\sigma}{\lambda(2-\lambda)}} \left[ 1- e^{-\lambda(2-\lambda)/\tau} \right].
\]
  
Let us now {reduce the size} of the triangles in Galton experiment, the shrinking parameter and the time frequency in such a way to still observe the evolution of the variance towards a {value bounded away from zero}. By looking at the coefficients of equation \fer{vava} we realize that this can be done by setting, for $\e \ll 1$
 \be\label{scala}
 {\sigma \to \e \sigma; \quad \lambda \to \e \lambda; \quad \tau \to \e\tau}.
 \ee
The meaning of the scaling \fer{scala} is clear.   Within this scaling, each {interaction produces only a small variation}. To see its effect on the distribution one has to suitably increase the time frequency of the interactions!

In this regime, for a given regular observable function $\varphi(x)$, expanding in Taylor series up to the second order, and neglecting terms of order $o(\e)$ (cf. Ref. \cite{FPTT} for details) we can write
  \[
  \langle \varphi(x_*) - \varphi(x)\rangle \simeq \varphi'(x) \langle x_* -x \rangle + \frac 12 \varphi''(x) \langle( x_* -x )^2\rangle \simeq
  \]
  \[
 {\e }\left[ - \lambda \, x \, \varphi'(x) + \frac \sigma 2\,   \varphi''(x) \right].  
  \]
  
 Setting for simplicity $\tau = {\e} $ we conclude that the kinetic equation is well approximated by
\be\label{fp}
\frac d{dt} \int_\R \varphi(x)f(x,t)\, dx = \int_\R \left[{ - \lambda\, x\, \varphi'(x) + \frac \sigma 2\,   \varphi''(x)} \right] f(x,t) \,dx.
 \ee
Equation \fer{fp} is the weak form of the Fokker--Planck equation \cite{FPTT}
\[
\frac{\partial f(x,t)}{\partial t} ={\frac\sigma 2}  \frac{\partial^2 f(x,t)}{\partial x^2} + {\lambda}\frac{\partial(x\, f(x,t))}{\partial x}.
 \]
The equilibrium solution of the Fokker--Planck equation (of unit mass) is the {Gaussian probability density }
\be\label{gd}
 f_\infty(x) = \sqrt{\frac {{\lambda}} {{4\pi \sigma}}} \exp \left\{ -\frac{\lambda x^2}\sigma\right\}.
 \ee
The kinetic approach justifies the {experimental evidence}. 
\begin{figure}
    \centering
    \includegraphics[width=0.7\textwidth]{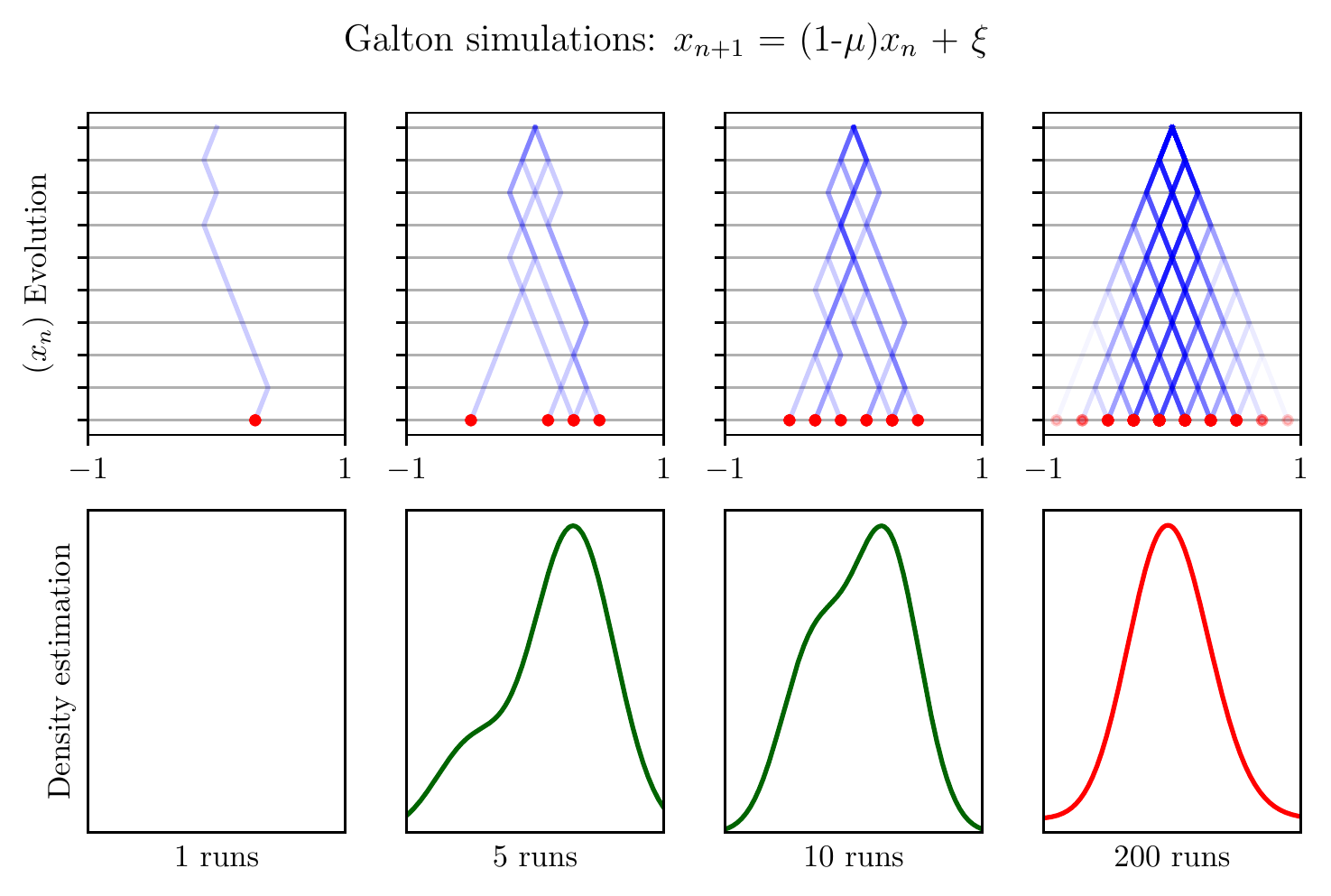}
    \caption{Trend of the modified Galton experiment with shrinked domain. }
    \label{fig:gn}
\end{figure}
\begin{remark}
At difference with the classical Galton bean machine, where, as explained above,  the normal density can be observed only if all the repeated trials start from $x=0$, in the kinetic approach, that leads to the Fokker--Planck equation, any initial probability density  of finite variance  is shown to converge towards the Gaussian density \fer{gd} \cite{FPTT}.
\end{remark}

\subsection{The lognormal distribution}

The same strategy can be applied to give a kinetic description of other experiments similar to Galton bean machine.  Few years later the contribution by Galton to demonstrate the central limit theorem, a similar device for the lognormal distribution was constructed by Kapteyn in Ref.   \cite{Kap}, while studying and popularizing the statistics of the lognormal in order to help visualize it and demonstrate its plausibility. A photograph of this machine is present in Ref.    \cite{AB}. 
The {elementary interaction} leading to the lognormal distribution is given by
  \be\label{ga-ln}
  x_* = x(1+ \sqrt\sigma\eta), 
  \ee
 where $x \ge 0$, and  $\eta$ is a {symmetric random variable} taking values $\pm 1$, and $0<\sigma < 1$, which guarantees $x_* \ge 0$.
 The elementary interaction \fer{ga-ln} can be easily written in the form of the elementary Galton interaction \fer{ga-in} by passing to the logarithmic scale. Indeed $y = \log x$ satisfies
 \[
  y_* = \log x_* = \log x + \log(1+ \sqrt\sigma\eta) = y  + \log(1 +\sqrt\sigma\eta).
 \]
As before, let us shrink the domain of $y$. For a given positive constant $\lambda < 1$ we consider the interaction
\[
  y_* =  y(1-\lambda) + \log(1 +\sqrt\sigma\eta).
\]
Going back to the original variable $x$ we obtain
\[
x_* = x^{1-\lambda} (1+\sqrt\sigma\eta),
\]
which can be fruitfully rewritten as
\be\label{s-int}
x_* = x - \lambda \frac{1 - x^{-\lambda}}\lambda \cdot x+ \sqrt\sigma x^{1-\lambda} \eta.
\ee
Note that, at difference with the elementary interaction \fer{ga-sc}, the coefficients of $x$ and $\eta$ are no more constant, but depend on $x$.  In particular, if we scale as before the parameters according to \fer{scala} we obtain 
\[
x_* =  x -\Phi^\e(x) x + \Psi^\e(x) x \eta,
\]
where
\[
\Phi^\e(x) =  \e \lambda \frac{1 - x^{-\e\lambda}}{\e\lambda}, \quad  \Psi^\e(x) = \sqrt{\e\sigma}x^{-\e\lambda}.
\]
Note that, for $x >0$
\be\label{graz1}
\lim_{\e \to 0} \frac 1\e \, \Phi^e (x) = \lambda \log x, \quad  \lim_{\e \to 0} \frac 1{\sqrt\e}\, \Psi^e (x) = \sqrt\sigma.
\ee
Hence, as shown in details in Refs.  \cite{GT1,GT2}, within the scaling \fer{scala} and small values of $\e$,  the  solution to the kinetic equation \fer{bo} is well approximated by the solution of the Fokker--Planck equation
 \be\label{FP-log}
  \frac{\partial f(x,t)}{\partial t}= \frac {\sigma} 2 \frac{\partial^2 }{\partial x^2}
 \left(x^2 f(x,t)\right ) +\lambda \frac{\partial}{\partial x}\left[\left(x \, \log x\right) f(x,t) \right].
 \ee
This time the steady state of unit mass of the Fokker--Planck equation \fer{FP-log} is the lognormal density
\be\label{log-n}
 f_\infty(x) = \sqrt{\frac\lambda{2\pi \sigma}}\frac 1x \exp \left\{ -\frac\lambda\sigma\left( \log x - \frac\sigma{2\lambda}\right)^2\right\}.
\ee
\begin{figure}
    \centering
    \includegraphics[width=0.7\textwidth]{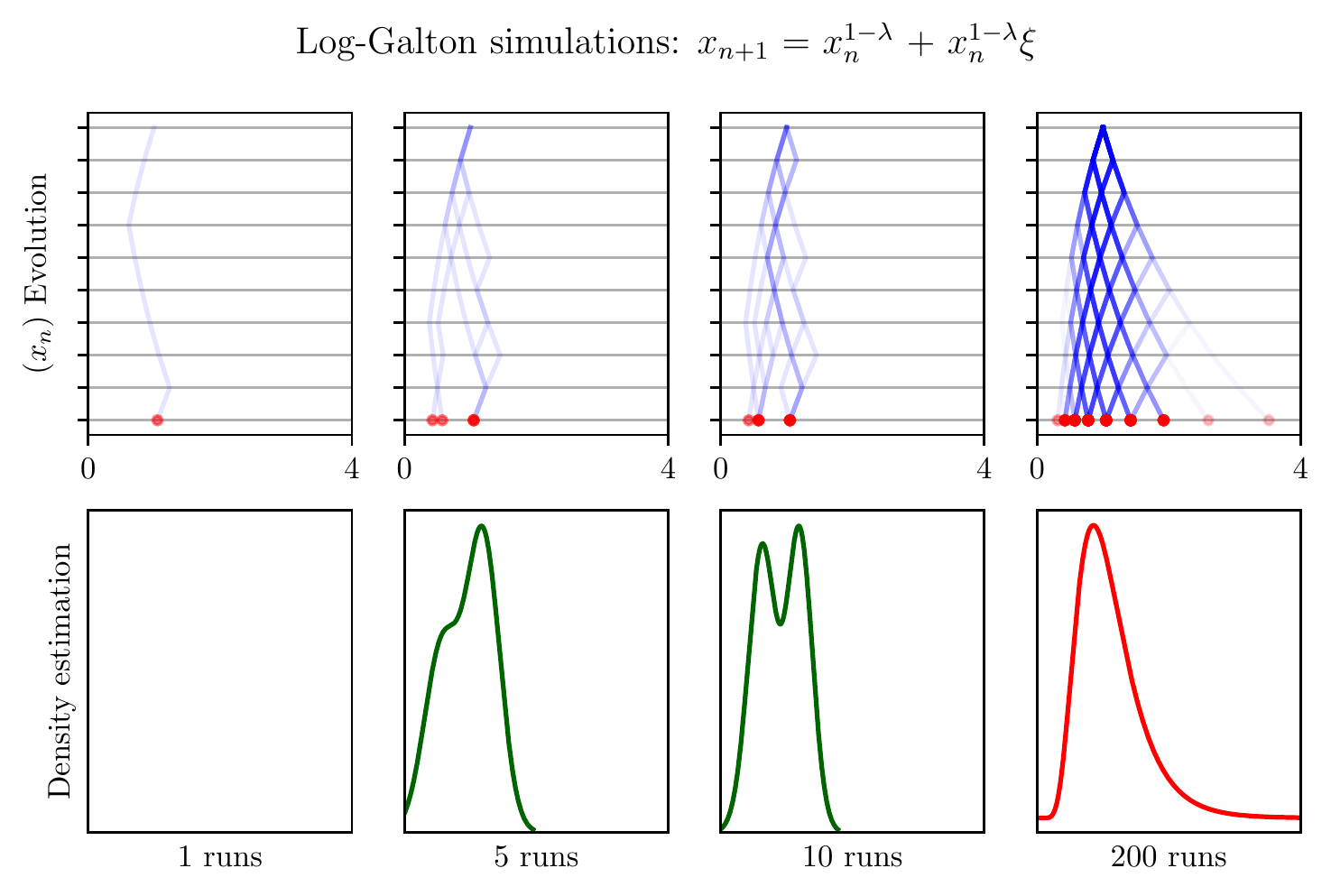}
    \caption{Trend of the modified Kapteyn experiment with shrinked domain. }
    \label{fig:ln}
\end{figure}

The kinetic description of the original experiments of Galton \cite{Gal1,Gal2},  and Kapteyn \cite{Kap} clarifies that most probability densities can be obtained by repeated trials in which each step is characterized by an elementary update of type \fer{coll1}, in which the old value is updated by a sum of two contributions, one deterministic and one random, which ultimately determine the shape of the self-similar or steady profile.  

\subsection{The Beta distribution}

While the two previous examples refer to densities taking values in $\R$ (Galton bean machine), and $\R_+$ (Kapteyn bean machine), the same strategy allows to obtain at the ground also densities supported in a bounded interval.
A simple example, which is directly connected to the problem treated in this paper, is concerned with the repeated trials which produce a Beta distribution. The mathematical problem which is at the origin of this elementary update is motivated by the study of the social problem of opinion formation \cite{Tos}. In this case
\be\label{opi}
x_* = x -\lambda(x-\mu) + \eta \sqrt{\sigma(1-x^2)}.
\ee
In \fer{opi}   $\eta$ is a {symmetric random variable} taking values $\pm 1$, and $0<\sigma, \lambda < 1$. Also, the constant $\mu\in(-1,1)$. Further, in \fer{opi} $\sigma$ and $\lambda$ are chosen in such a way to guarantee that $x_* \in (-1,1)$. At difference with the previous cases, the coefficient of the random variable $\eta$ decreases to zero as $x \to \pm 1$. The meaning of the elementary interaction \fer{opi} is clear. The opinion at each step is modified by two reasons. The first one is the compromise, represented by the deterministic part $\lambda(x-\mu)$. In absence of randomness, the distance of the personal opinion from the constant value $\mu$ (the mean opinion of the society) will decrease at each interaction. The second contribution is related to the self-thinking, namely to the possibility to change at random the personal opinion. This change is related to the distance from the extremal opinions $x = \pm 1$, and it is designed so that agents with opinions close to the extremal rarely change their opinions. 
If we rewrite interaction \fer{opi} in the form \fer{coll1} we have
\be\label{opi1}
x_* = x -\lambda\left(1-\frac\mu{x} \right) x+ \eta x\,\sqrt{\sigma\left(\frac1{x^2} -1\right) }.
\ee
In particular, if we scale as before the parameters according to \fer{scala} we obtain the elementary interaction
\[
x_* =  x -\Phi^\e(x) x + \Psi^\e(x) x \eta,
\]
where now
\[
\Phi^\e(x) =  \e \lambda\left(1-\frac\mu{x} \right) , \quad  \Psi^\e(x) = \sqrt{\e\sigma} \sqrt{\frac1{x^2} -1 }.
\]
In the grazing limit,  the  solution to the kinetic equation \fer{bo} is well approximated by the solution of the Fokker--Planck equation
 \be\label{FP-beta}
  \frac{\partial f(x,t)}{\partial t}= \frac {\sigma} 2 \frac{\partial^2 }{\partial x^2}
 \left(1- x^2) f(x,t)\right ) +\lambda \frac{\partial}{\partial x}\left((x-\mu) f(x,t) \right).
 \ee
By setting $\gamma = \sigma/\lambda$ one verifies that  the steady state of unit mass of the Fokker--Planck equation \fer{FP-log} is the Beta density $B(a,b)$ \cite{Tos}, where
\be\label{beta}
 a =  \frac{1-\mu}\gamma, \quad b= \frac{1+\mu}\gamma.
 \ee
 A symmetric Beta density is obtained by taking $\mu =0$. 
\begin{figure}
    \centering
    \includegraphics[width=0.7\textwidth]{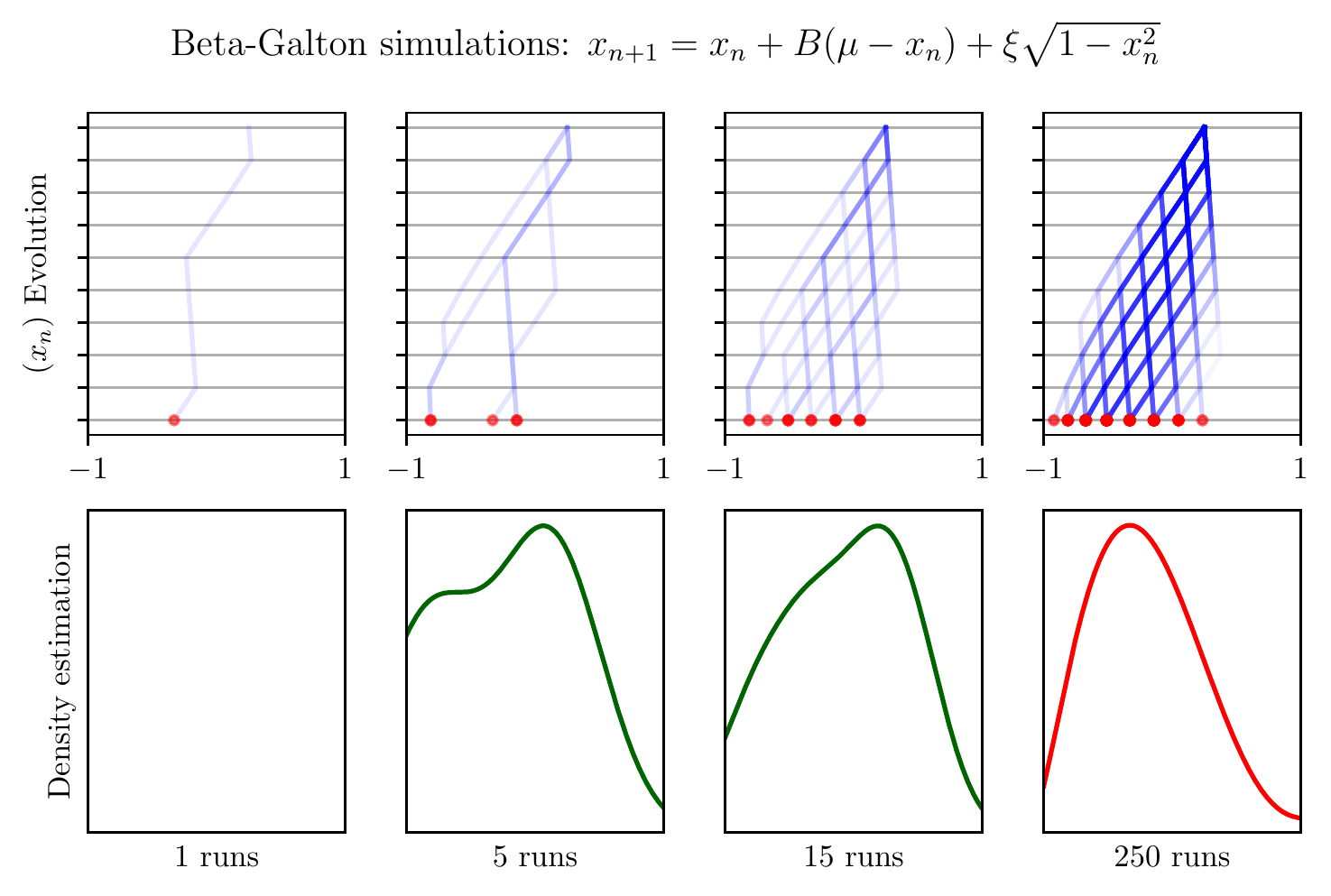}
    \caption{Beta trail experiment. }
    \label{fig:bn}
\end{figure}

\section{Kinetic modeling of body mass variations}\label{kinetic}

\subsection{The elementary update}\label{sec:update}

The goal of this Section is to characterize the properties of the stationary statistical distribution of weight  in a population  of growing organisms by  resorting  to the method briefly explained in Section \ref{sec:Galton}, namely by applying the approach of kinetic theory of multi-agent systems \cite{PT2}.
In what follows, we assume that the behavior of the population with respect to the weight growth is homogeneous. This homogeneity assumption is clearly quite strong in general, since it requires at least to restrict the population with respect to some characteristics, like age and sex.  

Once the homogeneity assumption is satisfied, the grower's state at any instant of time $t\ge 0$ is completely characterized by the value $w \ge0$ of  its weight.  We assume that this value is measured in terms of some unit. 
The unknown is the density (or distribution function) $f = f(w, t)$, where $w\in \R_+$ and the time $t\ge 0$, and the target is to study the statistical features of the subsequent steady state.
Without loss of generality, we assume that the density function is normalized to one
\[
\int_{\R_+} f(w, t)\, dw = 1.
\]
Then, for a given interval $A \subset \R_+$, the quantity
\[
\int_{A} f(w, t)\, dw.
\]
will denote the percentage of organisms with weight $ w \in A$. 

 The leading idea of kinetic theory is to express the dynamics of the distribution of a certain phenomenon in terms of the microscopic process ruling its elementary changes. 
At variance with the classical kinetic theory of rarefied gases, which aims at describing the dynamics of a huge number of particles subject to binary collisions,  the dynamics of population weight is governed by the changes due to the succession of generations,  where each generational change corresponds to an elementary update.

Thus, the temporal dynamics of the velocity distribution of collision-prone gas particles is identified with the temporal dynamics of the population weight distribution resulting from evolution related to the succession of new generations, and the trend at the equilibrium distribution of the velocity distribution of gas molecules \cite{Cer}, is identified with the statistical weight distribution observed at present.

The main problem in this approach is to correctly identify, from a mathematical point of view, the structure of the elementary update, which must be in agreement with both the biology of growth and the behavioral stimuli related to it.  
Similarly to the problems treated in Refs. \cite{DT,GT1,GT2}, the mechanism of weight growth can be postulated to depend on some universal features that can be summarized by saying that organisms are genetically programmed to increase the  value $w$ of their weight via nutrition, while manifesting resistance to decreasing it. 
Proceeding as in Ref. \cite{DT2},  we propose to model the elementary weight upgrade  in the form
 \be\label{true-coll}
 w^* = w  - \Phi^\e\left(\frac w {w_T}\right) w +  \Psi^\e(w)\, w\, \eta.
 \ee
In any new generation the value $w$ of the weight of the organism can be modified for two different reasons, both quantified by dimensionless coefficients acting on the actual weight variable $w$. The first one is  the (value) function $\Psi^\e(w/w_T)$, characterizing the asymmetric predictable behavior of growing organisms, which can assume assume both positive and negative values, and it is designed to respect at best the energetic and behavioral constraints. The second coefficient quantifies, in dependence of the energetic constraints measured by the function $\Psi^\e(w)$, the amount of unpredictability due to the presence of the environment. The uncertainty is contained in the random variable $\eta$,  negligible in the mean, and in any case are not so significant to produce a sensible variation of the value $\Psi^(w)\, w$.  Thus, with each new generation the weight of the organisms can both increase and decrease, and the mean value of this variation is fully determined by the function $\Phi^\e$. 
Last, the positive parameter $\e \ll1 $ quantifies the intensity of a single update.

The deterministic part is described by the  function \cite{DT2,PTZ}
 \be\label{vff}
 \Phi^\e(s) =  \mu \frac{1 - \exp\{\e(s^{-\delta} -1)/\delta\}}{1 + \exp\{\e(s^{-\delta} -1)/\delta\} } , \quad  s \ge 0,
 \ee
where $\mu$ and $\delta$ are positive constants, and $\delta \le 1$. Starting from $s=0$, for any fixed value of the  parameters $\e$ and $\delta$, the function $\Phi^\e(s)$ is convex in a small interval contained in the interval $(0,1)$, with an inflection point in $\bar s <1$, then concave.  The function in \fer{vff} is negative in the interval $s<1$ and positive in the interval $s >1$.

\begin{figure}
    \centering
    \includegraphics[width=0.7\textwidth]{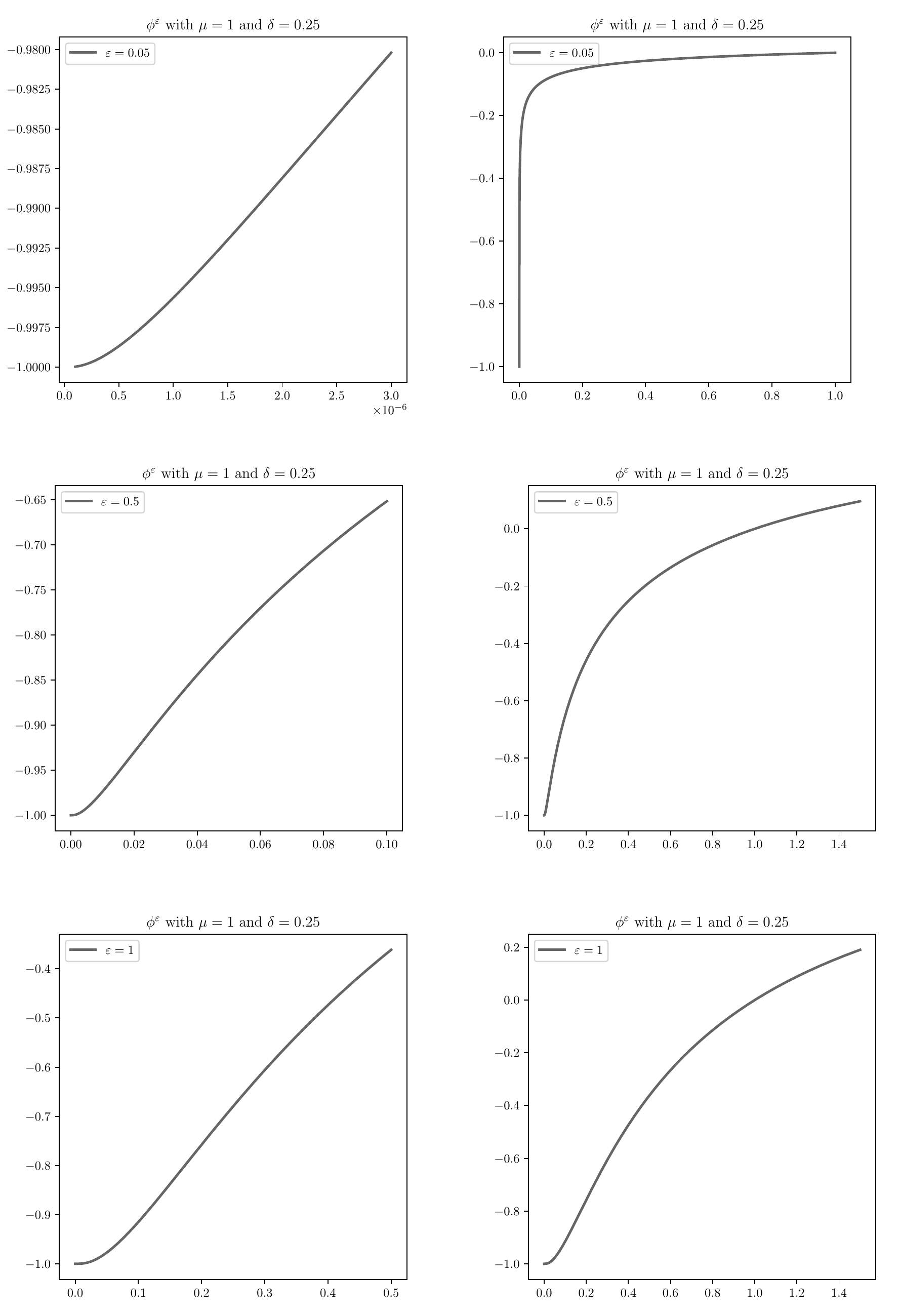}
    \caption{Trend of the function $\phi^\e$ for three values of $\e$. The other two parameters are fixed. The left-side of this plot is a zoom that allows to locate the exact position of inflection point. On the right side, one can see both the change in sign of the function, happening at $ s = 1$, and the asymptotic (concave) trend.}
    \label{fig:phieps}
\end{figure}

{Figure \ref{fig:phieps} reports the the trend of the function $\phi^\e$ for three different values of $\e$.
We note that the inflection point is closer to zero as $\e \to 0$}.

The value $w = w_T$, namely the optimal weight, corresponds to the reference point $s=1$, and $\Phi^\e(1) $ is equal to zero.  
 
  The value function  $\Phi^\e$ is bounded from below and above, and satisfies the bounds
 \be\label{bbb}
- \mu \le \Phi^\e(s) \le \mu \frac{1 - e^{-\e/\delta}}{1 +e^{-\e/\delta} }.
 \ee

 The lower and upper bounds in \fer{bbb} characterize the maximum amounts of change in body weight of a new generation over the previous one in terms of the positive constants $\e$, $\mu$, and $\delta$.  However, while the lower bound is independent of $\e$ and $\delta$, the upper bound is not. Moreover, for $\e\ll 1$ the upper bound is very close to zero, namely to the value of $\Phi^\e$ at the reference point $s=1$, which characterizes the optimal value $w = w_T$. 

  It is interesting to remark that, in the limit $\delta \to 0$  the  function \fer{vff} becomes
 \be\label{logn}
  \Phi_0^\e(s) = \mu \frac{s^\e -1}{s^\e +1} , \quad  s \ge 0,
 \ee
In this case the limit value function \fer{logn}, at difference with the value functions \fer{vff}, is concave, and the inflection point is lost.  This  value function was originally considered in Refs.  \cite{GT1,GT2}, where it was shown that it characterizes social phenomena well-represented by the lognormal distribution. 

According to the prospect theory of Kahneman and Twersky \cite{KT}, the function $\Phi^\e$ satisfies most of the properties of a \emph{value function}.
The notion of value function was originally related to various situations concerned with decision under risk. Kahneman and Twersky in  Ref. \cite{KT} identified the main properties characterizing a value function $\Phi(s)$, where $s \ge0$, in a certain behavior around the reference point $s=1$, expressed by the conditions
 \be\label{ccd}
 - \Phi\left(1-\Delta s \right) > \Phi\left(1+ \Delta s \right),
 \ee
 and 
 \be\label{cce}
 \Phi'\left(1+\Delta s \right)< \Phi'\left(1-\Delta s \right), 
 \ee
 where $\Delta s >0$ is such that $1- \Delta s \ge 0$.
 These properties are well defined for deviations from the reference point $s = 1$, and imply
that the value function below the reference point is steeper than the value function above it.
In other words, a value function is characterized by a certain asymmetry with respect to the
reference point $s = 1$. In the present situation, this asymmetry has a precise meaning. Given  two organisms with a weight at the same distance $\Delta s$ from the reference value $s=1$ from below and above, the grower's stimuli to reach the reference point are greater if it is below the reference point. It is easy to show that the function $\Phi^\e(s)$ in \fer{vff} satisfies properties \fer{ccd} and \fer{cce}. Note that these properties translate in a mathematical way that organisms are genetically programmed to increase the  value $w$ of their weight via nutrition, while manifesting resistance to decreasing it.

A third essential feature of the value function $\Phi^\e$ follows by considering its behavior as $\e \to 0$. Since
\be\label{dde}
 \lim_{\e \to 0} \frac 1\e {\Phi^\e\left( \frac w{w_T}\right)} = \frac\mu{2\delta} \left(1-\left( \frac{w_T}{w} \right) ^\delta\right),
 \ee
for small values of the parameter $\e$, namely for small variations of the weight in a single interaction, the deterministic part of the transition functions  \fer{true-coll}, as we shall detail in the forthcoming Section \ref{sec:growth}  can be put in close relation with the class of the  von Bertalanffy growth models \cite{Bert1,Bert2}. 

To conclude this Section, let us discuss the reasons leading to a correct mathematical modeling of the function $\Psi^\e(w)$, expressing the random fluctuation of the weight. This function should describe both the effect of the environment on random variations in weight, and the (limited) range over which the organism's weight can be influenced by the environment. In the case of a species characterized by flying, for example, the possibility of having an energy ratio favorable to flight in the presence of gravity is clearly limited to a superiorly limited weight range.
Among others, this behavior can be described by the function
\be\label{fluct}
\Psi^\e(w) = \frac{\sqrt\e}{\max\{ \sqrt\e, w^{\delta/2}\}} \left[\left(\frac p{1+\sqrt\e}w_T\right)^\delta -w^\delta) \right]^{1/2}_+.
\ee
In \fer{fluct} the positive constant $p$ is such that
\[
\frac p{1+\sqrt\e} >1,
\]
and, for a given function $h(w)$, $h(w)_+$ denotes the positive part of $h$, that is $h(w)_+= h(w)$ whenever $h(w) >0$, while $h(w)_+= 0$ if  $h(w) \le 0$. This choice implies that the deviation from a random weight fluctuation proportional to the weight itself decreases as the weight increases, becoming zero close to the critical threshold $p\,w_T$, located above the optimal weight $w_T$.

 The  function  $\Psi^\e$ is bounded from below and above, and satisfies the bounds
 \be\label{fff}
0\le \Psi^\e(s) \le (pw_T)^{\delta/2}.
 \ee
The presence of the function $\Psi_\e$  implies that the weight of the organisms can not exceed the value $pw_T$. Indeed, above the value $w_T$, the function  $\Phi^\e$ assumes negative values, and  the unique positive contribution to the growth comes from the random fluctuations.  For a given positive constant $\alpha\le 1$, let $ w$   by defined by
\be\label{bb}
 w = \frac {\alpha p}{1+\sqrt\e}\, w_T,
\ee
where the positive constant $\alpha$ is such that $(1+\sqrt\e)/p < \alpha < 1 +\sqrt\e$, and $w \ge w_T$. If $\alpha \in (1, 1 +\sqrt\e)$, $\Psi^\e(w) = 0$, the fluctuations are absent and the upper bound is respected.  In the remaining interval of $\alpha$  one has the upper bound
\[
\Psi^\e(w) =  {\sqrt\e} \left[\left(\frac{1}{\alpha}\right)^\delta -1 \right]^{1/2} \le {\sqrt\e} \left(p^\delta -1 \right)^{1/2}.
\]
The last inequality follows from \fer{bb}, that implies $\alpha \ge 1/p$. Therefore, if $ w \ge w_T$ 
  \[
 \Psi^\e( w) \eta \le  {\sqrt\e} \left(p^\delta -1 \right)^{1/2}\eta \le \sqrt\e
 \]
provided the random variable $\eta$ satisfies the upper bound
\be\label{up}
\eta \le \frac 1{(p^\delta -1)^{1/2}}.
\ee
Then \fer{true-coll} implies
\[
 w^* \le \bar w\, ( 1 + \Psi^\e( w) \eta) \le \frac {\alpha p}{1+\sqrt\e}\, w_T\left( 1 + \sqrt\e \right) \le p w_T.
\]
On the other side, the bounds \fer{bbb} and \fer{fff}  imply
 \[
 w^* \ge w\left( 1 -\mu + (pw_T)^{\delta/2} \eta\right).
 \]
Hence, in order that the weight $w^*\ge 0$, the  random variable $\eta$ must satisfy the condition 
 \be\label{down}
 \eta \ge -\frac{1-\mu}{(pw_T)^{\delta/2}}.
 \ee
If this is the case, the elementary update \fer{true-coll} maps non-negative values into non-negative values independently of the value of $\e$. Both the lower bond \fer{down} and the upper bound \fer{up} do not depend on the intensity $\e$ of the elementary upgrade \fer{true-coll}. 

However, it is important to remark that the previous bounds imply an upper bound on the variance $\sigma$ of the random variable $\eta$, namely a bound on the weight fluctuations. 

A further insight into the kinetic equation \fer{B1} can be obtained by looking at the evolution of the mean value of the weight driven by the elementary update \fer{true-coll}. To this aim, we compute the evolution of the observable $\varphi(w) = w$.  Thanks to \fer{dri}  we obtain
\begin{equation}\label{B1-mean}
\dfrac{d}{dt} \int_{\mathbb R_+} w f(w,t)dw = \frac 1\e \left \langle \int_{\mathbb R_+}  (w^*-w)f(w,t)dw \right\rangle 
= \int_{\mathbb R_+} \frac 1\e\, \Phi^\e\left(\frac w {w_T}\right) w \,f(w,t)dw.
\end{equation}
However, if $\e \ll 1$, in view of \fer{dde},  
\begin{equation}\label{eq:Phi_epsi}
\Phi^{\epsilon}\left( \frac w{w_T} \right) \approx \e \, \frac\mu{2\delta} \left(1-\left( \frac{w_T}w \right) ^\delta\right). 
\end{equation}
Therefore, 
\[
\dfrac{d}{dt} \int_{\mathbb R_+} w f(w,t)dw 
= \int_{\mathbb R_+} \frac 1\e\Phi^\e\left(\frac w {w_T}\right) w \,f(w,t)dw  \approx \frac\mu{2\delta} \int_{\mathbb R_+} \left(1-\left( \frac{w_T}w \right) ^\delta\right) w \,f(w,t)dw,
\]
namely an observable finite evolution of the mean value for any value of the parameter $\e$. 

 \subsection{Connections with classical growth models} \label{sec:growth}
 
 It is well recognized that body size provides an essential parameter in ecological and evolutionary research because of the ease with which it can be measured. It correlates with many ecological and life history traits of species, and thus provides a convenient surrogate for variables that can cause evolutionary and ecological patterns, but are more difficult to estimate.  Partly for this reason,  the identification of the growth law followed by body size frequency distributions for different organisms have long been topics of interest to biologists \cite{Black,Brown,Pet,Put}. The variation in the size of organisms and the dynamics of growth is nowadays explained by two major branches of theory (mechanistic and phenomenological).

A variety of mathematical models of growth, generally expressed by first-order ordinary differential equations, have been proposed under the names of Malthus, Verhulst, Gompertz, Richards, von Bertalanffy, West, and so on.

Most of the well-known models in the literature can be described by a unified version through a class of first-order differential equations of Bernoulli type for the variable $w(t)$, value of the weight of the organism
\be\label{gen}
 \frac{dw(t)}{dt} = \frac\alpha\gamma\, w(t) \left( 1 - \left(\frac{w(t)}{w_T}\right)^\gamma \right),
 \ee
parameterized by  $\alpha >0$, $\gamma\in[-1,1]$, and the  maximum weight limit  $w_T$.
Equation \fer{gen} include the logistic, von Bertalanffy and Gompertz growths. 
 If $\gamma \not=0$, equation \eqref{gen} can be easily integrated to get the analytical solution
 \be\label{solu}
 w(t) = w_T \left\{  \left[  \left(\frac{w_T}{w_0} \right)^\gamma  - 1  \right] e^{-\alpha t}  +1  \right\}^{-1/\gamma}.
 \ee
describing the growth of the weight of the organism starting from the initial value $w_0$ at time $t = 0$ toward the stable equilibrium represented by the maximum weight limit $w_T$.

In the context of biological growth processes,  growth models of type \fer{gen}, characterized by negative values of the constant $\gamma$, seem to be more in keeping with reality \cite{Bert1,Bert2,West}.  The most known model in this range of the parameter is due to von Bertalanffy,  and it is usually written as
\begin{equation}\label{eq:vonB}
\frac{dw(t)}{dt} = p w(t)^a - qw(t), 
\end{equation}
where $0\le a<1$, and  $p,q>0$ are the rates of growth and size-proportional catabolism, respectively. This model is easily obtained from \fer{gen} by assuming $\gamma = a-1 <0$, $\alpha = q(1-a)$ and  $w_T= \left({p}/{q} \right)^{{1}/{(1-a)}}$. 
The limit case $\gamma \to 0$ in \fer{gen} corresponds to Gompertz growth. 

These models of growth give rise to similar evolutions, as expected, of course, because they need to fit the same experimental trends. The literature on the subject is huge. So, for more information we refer to the recent review paper \cite{Mar} and to the article \cite{Rick}.

Despite the apparent simplicity of the question, there is currently no general consensus on the type of growth law that is best used to fit the data, which is why randomness often plays an important role.  Growth in a random environment has been formulated in the context of stochastic birth and death processes by several authors (see, e.g., Refs. \cite{Nobile,Ricciardi,Tan} and their exhaustive bibliographies) to account for environmental fluctuations. 

 It important to remark that, for small values of the parameter $\e$, namely for small variations of the weight in a single update, the deterministic part of the elementary upgrade  \fer{true-coll}  is closely related with the class of growth equations \fer{eq:vonB}. 

In this case, expanding $\varphi(w^*)$ in Taylor's series at the order one we get
\[
\varphi(w^*) -\varphi(w) = \varphi'(w)(w^* -w) + R_\e(w),
\]
where the remainder $R_\e$ is  $o(\e)$ as $\e \to 0$.  Relation \fer{dde} then implies that the kinetic equation \fer{B1} takes the form
\begin{equation}\label{B2}
\dfrac{d}{dt} \int_{\mathbb R_+} \varphi(w) f(w,t)dw =  - \int_{\mathbb R_+}\varphi'(w)\frac\mu{2\delta}\, w\left(1-\left( \frac{w_T}w \right) ^\delta\right) 
  f(w,t)dw +\mathcal R_\e(t),
\end{equation}
where the dependence on the small parameter $\e$ is confined in the remainder term $\mathcal R_\e(t)$, which for $\e \ll 1$ vanishes as $\e \to 0$. Hence, taking the limit as $\e \to 0$ one shows that the remainder term vanishes, and integrating by parts the right-hand side of equation \fer{B2} it follows easily that equation \fer{B2}  is the weak form of the drift equation 
\be\label{con5}
 \frac{\partial f(w,t)}{\partial t} =  \frac\mu{2\delta}\,\frac{\partial}{\partial w} \left[w\left(1-\left( \frac{w_T}w \right) ^\delta\right)f(w,t)\right] .
  \ee
It is remarkable that also the deterministic dynamics of growth driven by equation \fer{gen}  leads to partial differential equations of type \fer{con5} which describe the evolution of the density function $f(w,t)$ which, at a certain time $t=t_0$, measures the statistics of the weight $w \ge 0$ in the population. 

Following the analysis of Ref. \cite{PTZ}, let $W(t)$  denote the process which describes the statistical distribution of the weight in the population at time  $t \ge 0$, and let $F(w,t)$ denote its probability distribution, defined by 
 \[
 F(w,t) = P(W(t) \le w), \qquad w \ge 0.
 \]
 
The classical way to recover the evolution of $F(w,t)$ consequent to a growth driven by equation \fer{gen} is to remark that, if $w(t)$ denotes the solution \fer{solu} to equation \fer{gen} departing from the value $w_0$ at time $t=0$, then 
 \[
 P(W(t) \le w(t)) = P(W(t=0) \le w_0),
 \]
or, what is the same
 \be\label{con2}
 F(w(t),t) = F(w_0,t=0) = const.
 \ee
Therefore, taking the time derivative on both sides of \fer{con2} one has
 \be\label{con3}
 \left.\frac d{dt} F(w(t),t) = \frac{\partial F(w,t)}{\partial t} + \dot w(t)\frac{\partial F(w,t)}{\partial w}\right|_{w =w(t)} = 0.
 \ee
Hence, using \fer{gen} into \fer{con3} it follows that $F(w,t)$  satisfies the conservation law
\be\label{con4}
 \frac{\partial F(w,t)}{\partial t} +\frac\alpha\gamma\,w\, \left( 1 - \left(\frac{w}{w_T}\right)^\gamma \right)\frac{\partial F(w,t)}{\partial w}= 0.
 \ee
If $F(w,t)$ is regular with respect to $w$, and  $f(w,t)$ denotes its derivative with respect to $w$, namely the probability density of the process $W(t)$,  setting  $\gamma = -\delta$ and taking the derivative with respect to $w$ on both sides of equation \fer{con4} one obtains that, in terms of the probability density $f(w,t)$ equation \fer{con4} is rewritten as \fer{con5}, where the constant $\mu/2$ is substituted by the constant $\alpha$.  

\begin{remark} The previous analysis clarifies that the time evolution of the statistical distribution of weight determined by the growth models \fer{gen} is coherent with the time evolution of the  \emph{grazing} limit of the kinetic model \fer{B1}, in absence of random fluctuations. However, the description of weight variations of organisms in terms of the elementary upgrade \fer{true-coll} is more general, and allows to better clarify the meaning of the parameters characterizing the value function $\Phi^\e$. 
\end{remark}

\section{The Fokker--Planck description}\label{Fokker}

\subsection{The limit of grazing interactions}\label{sec:grazing}

The discussion of Section \ref{sec:growth}, relative to the connection of the deterministic interaction with growth models, can be fruitfully extended to the general interaction \fer{true-coll} to recover the law of variation in the \emph{grazing} limit regime. Let us briefly describe this argument,  which can be seen in detail in Refs. \cite{FPTT,PTZ}. 

Let us consider again the weak form of the kinetic model \fer{B1} with the right time scale
\begin{equation}\label{eq:boltz2}
\dfrac{d}{d t}  \int_{\mathbb R_+} \varphi(w)  f(w,{t})\, dw = \dfrac{1}{\epsilon}\left \langle \int_{\mathbb R_+} (\varphi(w^*)-\varphi(w))  f(w, t)\, dw \right\rangle.
\end{equation}
If $\epsilon \ll 1$ is sufficiently small, the difference $w^* -w$ is small, and assuming $\varphi$ enough regular (at least $\varphi \in \mathcal C^3_0(\mathbb R_+)$)  we can expand it in Taylor series up to the order three to get
\[
\varphi(w^*) -\varphi(w) = (w^* -w)\partial_w \varphi(w) + 
\dfrac{1}{2} (w^* -w)^2\partial_w^2\varphi(w) + \dfrac{1}{6}(w^*-w)^3 \partial_w^3 \varphi(\bar w), 
\]
being $\bar w\in \left( \min\{w,w^*\},\max\{w,w^*\} \right)$. Substituting to the expression of $w^* -w$ its value obtained from \eqref{true-coll} and plugging it into the above expansion, from \eqref{eq:boltz2} we obtain
\[
\begin{split}
&\dfrac{d}{d{t}} \int_{\mathbb R_+} \varphi(w) f(w,{t})\, dw  = \dfrac{1}{\epsilon} \left[\int_{\mathbb R_+}  \Phi^\e(w/w_T)\,w \partial_w \varphi(w) f(w,{t}) \,dw + \right.\\
&\left. \dfrac{\sigma}{2} \int_{\mathbb R_+}\partial_w^2 \varphi(w) (\Psi^\e(w)\,w)^2 f(w,{t})\,dw \right] 
+ R_\varphi(f)(w,{t}),
\end{split}\]
where $R_\varphi(f)$ is the remainder
\[
\begin{split}
R_\varphi(f)(w,{t}) &=  \dfrac{1}{2\epsilon} \int_{\mathbb R_+}\partial_w^2 \varphi(w)\left(\Phi^\e(w/w_T)\right)^2w^2f(w,{t})\, \,dw \\
&+ \dfrac{1}{6\epsilon} \left\langle \int_{\mathbb R_+} \partial_w^3\varphi(\bar w) \left( \Phi^\e(w/w_T)\, w + \Psi^\e(w)\,w \eta \right)^3 f(w,{t})\,dw \right\rangle. 
\end{split}\]
Thanks to the assumed smoothness we argue that $\varphi$ and its derivatives are bounded in $\mathbb R_+$. Further, if $\eta$ has bounded moment of order three, namely $\langle |\eta|^3 \rangle <+\infty$, and observing that for $\epsilon\ll 1$ the transition function $\Phi^\e$ behaves like in \eqref{eq:Phi_epsi},
we can easily argue that in the limit $\epsilon \rightarrow 0^+$ we have
\[
\left| R_\varphi(f) \right| \rightarrow 0, 
\]

Hence, in the limit $\epsilon \rightarrow 0^+$ equation \eqref{eq:boltz2} converges to
\begin{equations}\label{FP3}
&\dfrac{d}{d{t}} \int_{\mathbb R_+} \varphi(w) f(w,{t})\,dw =\\
 &\int_{\mathbb R_+}  \frac\mu{2\delta}\left(1-\left(\dfrac{w_T}{w}\right)^\delta \right) w  f(w,{t}) \partial_w\varphi(w) \,dw + \dfrac{\sigma}{2} \int_{\mathbb R_+} w^{2-\delta}((pw_T)^\delta - w^\delta)  f(w,{t}) \partial_w^2\varphi(w)\,dw.
 \end{equations}
 Resorting to the discussion of the last part of Section \ref{sec:update}, it follows that the values of the weight variable $w$ are confined to the finite interval $(0, pw_T)$.
 
 Next, integrating back by parts we conclude that the limit density $f=f(w,t)$ is solution of the  Fokker-Planck equation 
 \be\label{FP-f}
  \frac{\partial f(w,t)}{\partial t}= \frac {\sigma} 2 \frac{\partial^2 }{\partial w^2}
 \left[w^{2-\delta}((pw_T)^\delta - w^\delta) f(w,t)\right ]+ \frac\mu{2\delta} \frac{\partial}{\partial w}\left[\left(1-\left(\dfrac{w}{w_T}\right)^\delta \right) w \, f(w,t) \right].
 \ee

Clearly, integration by parts is justified provided the following (no flux) boundary conditions are satisfied for all ${t}>0$
\be
\dfrac{\mu}{2\delta} \left( \left( \dfrac{w}{w_T} \right)^\delta -1\right) w f(w,{t}) +\dfrac{\sigma}{2} \partial_w \left[w^{2-\delta}((pw_T)^\delta - w^\delta) f(w,t)\right]\Bigg|_{w = 0}^{w=pw_T} =0 
\ee
The Fokker--Planck equation \fer{FP-f} retains memory of the kinetic description through the relevant parameters of the transition function \fer{true-coll}, namely the parameters $\delta$ and $\mu$, and through the random fluctuation. However,  the details of the variable $\eta$ are lost in the limit passage, so that the role of fluctuations is taken into account only through their variance, parameterized by $\sigma$.  As we shall see, at difference with the others, the  value of the parameter $\delta$ fully characterizes the shape of the steady state of equation \fer{FP-f}. 
 
 \subsection{Steady states are generalized
 Beta distributions}\label{sec:equilibria}

Equation \fer{FP-f} belongs to the class of  Fokker--Planck equations characterized by a variable coefficient of diffusion and drift which are suitable to describe social phenomena \cite{FPTT}, usually written in divergence form as
\be\label{FPnorm}
  \frac{\partial f(w,t)}{\partial t}=   \frac{\partial }{\partial w}\left\{   \frac{\partial}{\partial w}\left[\kappa(w)f(w,t)\right] + \nu(w)f(w,t)\right\} , \quad w \in (0, pw_T).
\ee
Then, the  equilibrium distribution $f_\infty(w)$ is obtained by imposing that the flux is vanishing, that is by looking for the solution of the first order differential equation
\be\label{eq}
 \frac{d}{dw}\left[\kappa(w)f(w)\right] + \nu(w)f(w) = 0.
\ee
\paragraph{Step 1} In the case under study
 \[
  \kappa(w) = \frac\sigma 2w^{2-\delta}((pw_T)^\delta - w^\delta); \quad 
 \nu(w) =  \frac\mu{2\delta}\left(1-\left(\dfrac{w_T}{w}\right)^\delta \right)\, w.
   \]
Let us set $v = w/w_T$. Then 
\be\label{coeff}
 \kappa(w) = w_T^2\, \frac\sigma 2  v^{2-\delta}(p^\delta - v^\delta) =w_T^2\, \tilde\kappa(v) ; \quad 
 \nu(w) =  w_T\,\frac\mu{2\delta} \left(1-v^{-\delta} \right)\, v = w_T\, \tilde\nu(v).
\ee
Also, since
\[
\frac{\partial}{\partial w} = \frac 1{w_T} \frac{\partial}{\partial v},
\]
by setting $g(v) = f(w)$ we obtain that, if $f(w)$  satisfies \fer{eq}, $g(v)$ satisfies the first order differential equation
\be\label{eq2}
 \frac{d}{d v}\left[\tilde\kappa(v)g(v)\right] + \tilde\nu(v)g(v) = 0,
\ee
that does not depend on the value $w_T$ of the optimal weight. 

\paragraph{Step 2}
Recalling expressions \fer{coeff},  $g(v)$, $v \in(0,p)$, solves
\[
 \frac{d}{d v}\left[ v^{2-\delta}(p^\delta - v^\delta) g(v)\right]  + \frac 1\lambda v^{1-\delta} \left(v^{\delta}-1 \right) g(v)=0,
\]
where we defined
\be\label{hetero}
\lambda =  \frac{\sigma\delta}\mu.
\ee
The positive constant $\lambda$, in reason of its proportionality to the variance $\sigma$ of the random fluctuations, is a measure of the \emph{heterogeneity} of the population of organisms.

Expanding  the derivative of the product,  and grouping similar terms we obtain the equivalent differential equation
 \[
 v^{2-\delta}(p^\delta - v^\delta) \frac{dg(v)}{d v} = v^{1-\delta} \left[ \left(\frac 1\lambda - (2-\delta) p^\delta \right) - \left(\frac 1\lambda - 2 \right) v^\delta  \right] g(v).
 \]
If $v >0$, we can divide both sides by $v^{1-\delta}$ to get
\be\label{ok2}
 v(p^\delta - v^\delta) \frac{dg(v)}{d v} = \left[ \left(\frac 1\lambda - (2-\delta) p^\delta \right) - \left(\frac 1\lambda - 2 \right) v^\delta  \right] g(v).
 \ee
 
\paragraph{Step 3}

Let us set $v^\delta = u$, and $g(v) = h(u)$.  Then
\[
v\frac{dg(v)}{dv} =v \frac{dh(u)}{du} \delta v^{\delta -1} = \delta u \frac{dh(u)}{du}. 
\]
Therefore, in terms of the function $h(u)$, equation \fer{ok2} takes the form
\be\label{ok3}
 u\, (A - u) \frac{dh(u)}{d u} =  \left( B  - C u  \right) h(u).
\ee
 where we defined
 \be\label{coo}
 A=p^\delta; \quad B= \frac 1\delta \left[\frac 1\lambda- (2-\delta) p^\delta \right] ; \quad C=  \frac 1\delta\left[\frac 1\lambda - 2 \right].
 \ee

 \begin{remark}
The signs of the coefficients $B$ and $C$ on the right-hand side of equation \fer{ok3} depend on the values of the parameters characterizing the deterministic and random parts of the elementary interaction \fer{true-coll}. In what follows, we require that these coefficients are positive. This is always possible simply by assuming that the heterogeneity $\lambda$  of the population is sufficiently small. 
\end{remark} 

Equation \fer{ok3} is solvable exactly. Indeed, it can be rewritten as
 \[
 \frac d{d u}\log h(u) = \frac{B-Cu}{u(A-u)} = \frac d{du}\log\left[u^{B/A}(A-u)^{C-B/A}\right].
 \]
 Hence, the unique solution of the differential equation \fer{ok3} of unit mass is a Beta distribution $B(a,b)$, supported on the interval $(0, p^\delta)$,  parameterized by the positive shape parameters
 \be\label{be}
 a= \frac 1\delta\left[ \frac 1{\lambda\,p^\delta} - 2(1-\delta)\right], \quad 
 b = \frac 1{\delta\lambda}\frac{p^\delta-1}{p^\delta}.
 \ee
Going back to the original differential equation \fer{eq} we conclude that the unique stationary solution of the Fokker--Planck equation \fer{FP-f} is expressed by 
 \be\label{equi}
 f_\infty (w) = D\,{ \left(\frac w{w_T}\right)^{\delta (a-1)}\left[p^\delta - \left(\frac w{w_T}\right)^{\delta}\right]^{b-1} }
 \ee
supported on the interval $(0, pw_T)$, with $a$ and $b$  defined by \fer{be}. 
If we assume that $f_\infty$ has unit mass, so that
\[
\int_0^{pw_t} f_\infty(w) \, dw = 1,
\]
we can rewrite it in the  form
\be\label{gen-beta} 
 f_\infty (w) = \frac\delta{pw_T B(\alpha,\beta)}\,{ \left(\frac w{pw_T}\right)^{\delta\alpha-1}\left[1- \left(\frac w{pw_T}\right)^{\delta}\right]^{\beta-1} }
\ee
where $B(x,y)$ is the Beta function of parameters $x,y$, and
\be\label{new-be}
\alpha = a+ \frac 1\delta -1; \quad \beta =b,
\ee
and $a,b$ are defined in \fer{be}. 

Hence, the equilibrium solution of the Fokker--Planck equation \fer{FP-f}  is a Generalized Beta distribution of the first type (GB1 for short \cite{MX}).
\begin{remark}\label{r1} It is interesting to note that, provided the random variable $X$, taking values on the interval $(0,pw_T)$,  is distributed according to \fer{gen-beta}, the random variable $Y= (X/(pw_T))^{\delta}$ is Beta distributed on the interval $(0,1)$ according to $B(\alpha,\beta)$.
\end{remark}

\section{Numerical fitting}\label{fitting}
\subsection{Fitting Chiroptera populations} 
%\textcolor{red}{\textbf{Eleonora: }}
%
 In this section, we validate our kinetic model by fitting the actual statistical distribution of weights in a population of Chiroptera.
We use the dataset in Ref. \cite{dataset}, that contains both the weight and length of mammals living in the Atlantic zone.
As stated by the authors of Ref. \cite{dataset}, it is important to record trait variation not only \textit{between} species but also \textit{within} species.
From this point of view, mammals are interesting as they widely interact with both the environment (for instance, through pollination and seed dispersal) and other mammals (for instance through predation and grazing).
As remarked in Ref. \cite{blnick2011intraspecific}, variations among individuals of the same species is a key focus of evolutionary theory.

The dataset we consider contains a record for $39\,850$ mammals living in tropical and subtropical forests in Brazil, Paraguay, and Argentina.
For most records, the dataset gives the body mass, age, sex, reproductive stage, status of the animal and geographic coordinates of the \emph{samples}; for a view records, some data are missing.
In the dataset, we have nine different orders, each containing various species: Rodentia, Didelphimorphia, Carnivora, Primates, Cingulata, Artiodactyla, Pilosa, Perissodactyla, Chiroptera, Lagomorpha.
In this work, we focus on Chiroptera, for which we have weight data for $23\,010$ samples coming from 98 species (86.7\% of the known species of Chiroptera).
For our analysis, we only consider adult specimens of Chiroptera, for a total of $19\,612$ samples.
From now on, let $n$ be the number of samples of adult Chiroptera for which we have weight recorded, namely $n = 19\,612 $ 

Such specimens present a range that spans from 1 g (\emph{Peropteryx kappleri}) to 131 g (\emph{Chrotopterus auritus}).

According to Remark \ref{r1}, in what follows we prepare the dataset of weights first by normalizing them to lie in the unit interval, and second by changing each normalized weight $x$ in the data set to $x^\delta$. Finally, we fit the well-prepared data set with a Beta distribution to understand if they consistent with the theoretical discussion of the previous Section \cite{fitdistrplus}.
The motivation is twofold:
First of all, the Beta distribution is well known  in the communities of statisticians and ecologists, and thus,  readers are facilitated in understanding the results regardless of his or her field of research.
In addition, even though the Beta distribution has fewer parameters than the $GB1$, those parameters are sufficient to retrieve the information we need for validating our analysis.

\begin{remark}\label{r2}
Importantly, once we have obtained the optimal fit of the data with a Beta distribution, we know the constant values of the two parameters that characterize the Beta distribution. However, looking at the expressions of the parameters $\alpha$ and $\beta$, given by \fer{new-be}, we realize that they depend on the triplet $\delta, \lambda, p$, i.e., the growth parameter $\delta$, the heterogeneity parameter $\lambda$, and the parameter $p$ that characterizes the maximum size allowed for the flight of Chiroptera. Consequently, since the number of unknown parameters is higher than two,  the fitting is not enough to quantify their precise values. Therefore, we fix the value of the growth rate $\delta$ from the outset, following the analysis of West et al. in Ref. \cite{West}, relative to a general quantitative model, based on fundamental principles for metabolic energy allocation between maintenance of existing tissues and production of new biomass, in which a single universal parameterless curve emerged describing the growth of many different species. According to West's ontogenetic growth model in Ref. \cite{West}, and the discussion in Section \ref{sec:growth} on the connection between the elementary update \fer{true-coll} and growth models, we assume the value $\delta=1/4$. Furthermore, the growth rate predicted by the analysis in Ref.  \cite{West} allows us to identify the value of the constant $\mu/(2\delta)$ that characterizes the rate of time at which growth evolves (the parameter $a$ in Ref. \cite{West}). 
\end{remark}
 
\subsection{Calibrating the parameters}
We fit the dataset in accordance with the previous discussion.

For our analysis, we only consider adult specimens of Chiroptera, for a total of $19\,612$ samples.
To fix  notations, given $\hat{\bm{w}}$ the vector of data, we define $\hat{\bm{y}}$ the vector of the data raised to the  power $\delta =1/4$. 

The exact re-scaled $i$th-entry of the vector $\hat{\bm{y}}$ is obtained by 
\[
\tilde{y}_i = \dfrac{\hat{y}_i - \min{(\hat{\bm{y}})}}{\max(\hat{\bm{y}}) - \min(\hat{\bm{y}})}.
\]
It is immediate to observe that, from a numerical perspective, this definition is not working properly.
The reason is that $\hat{\bm{y}}$ might contain  the values $0$ and/or $1$, and this would lead to troubles during the computation of the log-likelihood for the Beta distribution. 
A simple solution to this problem is to add in the scaling two positive penalties $\varepsilon_1, \varepsilon_2$, with $\varepsilon_1 < \varepsilon_2$,  to obtain
\[
\tilde{y}_i = \dfrac{\hat{y}_i - \min({\hat{\bm{y}})} + \varepsilon_1}{\max(\hat{\bm{y}}) - \min(\hat{\bm{y}}) + \varepsilon_2}. 
\]
In our computation, we set $\varepsilon_1 = 10^{-7}$ and $\varepsilon_2 = 2\cdot 10^{-7}$.
In reason of this simple correction, the entries of $\tilde{\bm{y}}$ lie inside the open interval (0,1).

We can now fit the vector $\tilde{\bm{y}}$ with a standard Beta distribution with density function
\begin{equation}
    f^{(B)}_\infty(y) = \dfrac{y^{(1-\alpha)}(1-y)^{(1-\beta)}}{B(\alpha,\beta)}. 
    \label{eq:fitted}
\end{equation}
The resulting fitting returns the values
\be\label{true}
\alpha = 3.999092 \qquad \beta= 3.708544.
\ee

To validate our fitting, we propose two different strategies.
Firstly, from a numerical perspective, we compute two scores, namely the Akaike Information Criterion (AIC) and the  Bayesian Information Criterion (BIC) as scores related to the goodness of the fitting.
According to Ref. \cite{hastie2009theelements}, AIC and BIC are both minimized during the MLE optimization procedure, and thus, one may select the models with the lowest AIC and BIC.

In particular such values can be computed as
\[\text{AIC} \, = \, 2k - 2\ln(\hat L) \qquad \text{BIC} = k\ln(n) - 2\ln(\widehat L). \]
Where $\widehat L$ is the maximized value of the likelihood function, $n$ is the number of data points in the dataset, and $k$ is the number of parameters estimated by the model. 

In our fitting procedure, with all the above-mentioned assumptions, we get
\[\text{AIC} = -14503.71, \qquad \text{BIC} = -14487.94, \]
Note that, for this kind of analysis, a commonly chosen distribution is the lognormal \cite{GT1,GT2}.
In our case, if we fit vector data $\bm{\tilde{y}}$ with lognormal distribution, we get AIC = $-12267.28$ and BIC = $-12267.28$, which are clearly worse.
Better results are obtained using the Weibull distribution (AIC = $-14432.79$ and BIC = $-14417.02$).
Note that, to make such fittings comparable among them, we fit the vector $\tilde{\bm{y}}$ on all the three distributions.

\begin{figure}
    \centering
    \includegraphics[width=0.7\textwidth]{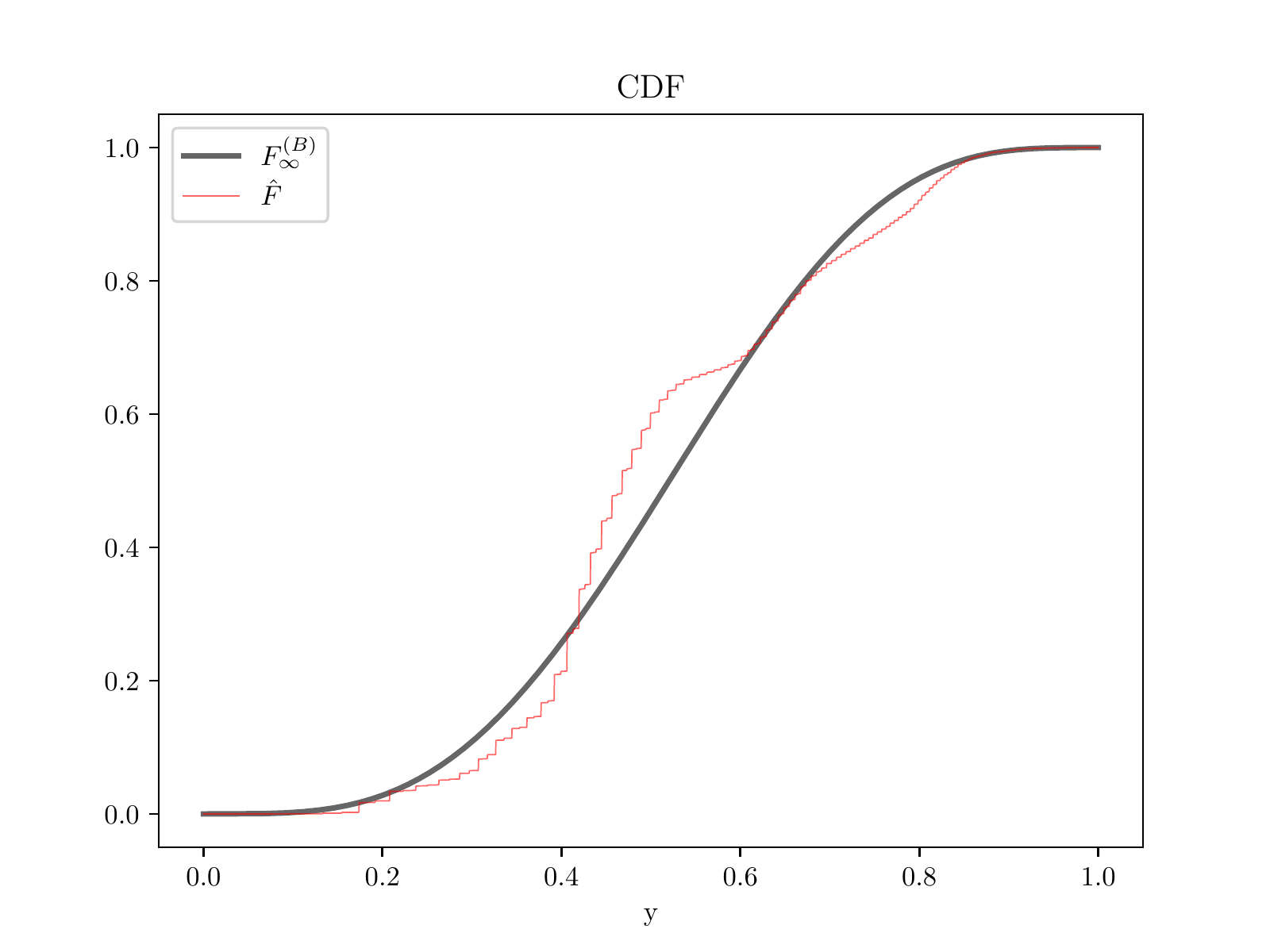}
    \caption{Plots of the estimated cumulative density function $\hat{F}$ (red), and $F^{(B)}_\infty$ (black) as derived from the fitted parameters.}
    \label{fig:cdf}
\end{figure}

Then, we visually analyze the trend of the Survival Function in log-scale, both analytically and computationally.
Let $F^{(B)}_\infty$ the Cumulative Distribution Function (CDF) of the Beta density \eqref{eq:fitted}.
Let $\hat{F}$ be the \emph{empirical} CDF.
This function can be computed by counting, for each $y \in [0,1]$, the number of observation $\tilde{y}_i$ that are less or equal than $y$, divided by the number of observation $n$: \[\hat{F}(y) = \frac1 n\sum_{i=1}^n \mathbf{1}_{[\tilde{y}_i \le y]}.\]
The values can be easily obtained, for instance, using the method {\ttfamily ecdf} in {\ttfamily R}.
Figure \ref{fig:cdf} reports a comparison between the two curves.

Giving any cumulative distribution function $F$, we can compute the   Survival Function (SF), defined as $S(y) =  1-F(y)$.
We plot SFs in semilog $y$-scale: 
this is a standard convention dating back to the work of Pareto \cite{pareto1897cours}, who observed that power laws, when looking at the SF in log-log scales, are characterized by a polynomial decay.
Here, due to the scaling between 0 and 1, we only plot the log on the $y$ axis, as the $x$-axis is subjected to numerical errors due to the presence of the values $0$ and $1$.
\begin{figure}
    \centering
    \includegraphics[width=0.7\textwidth]{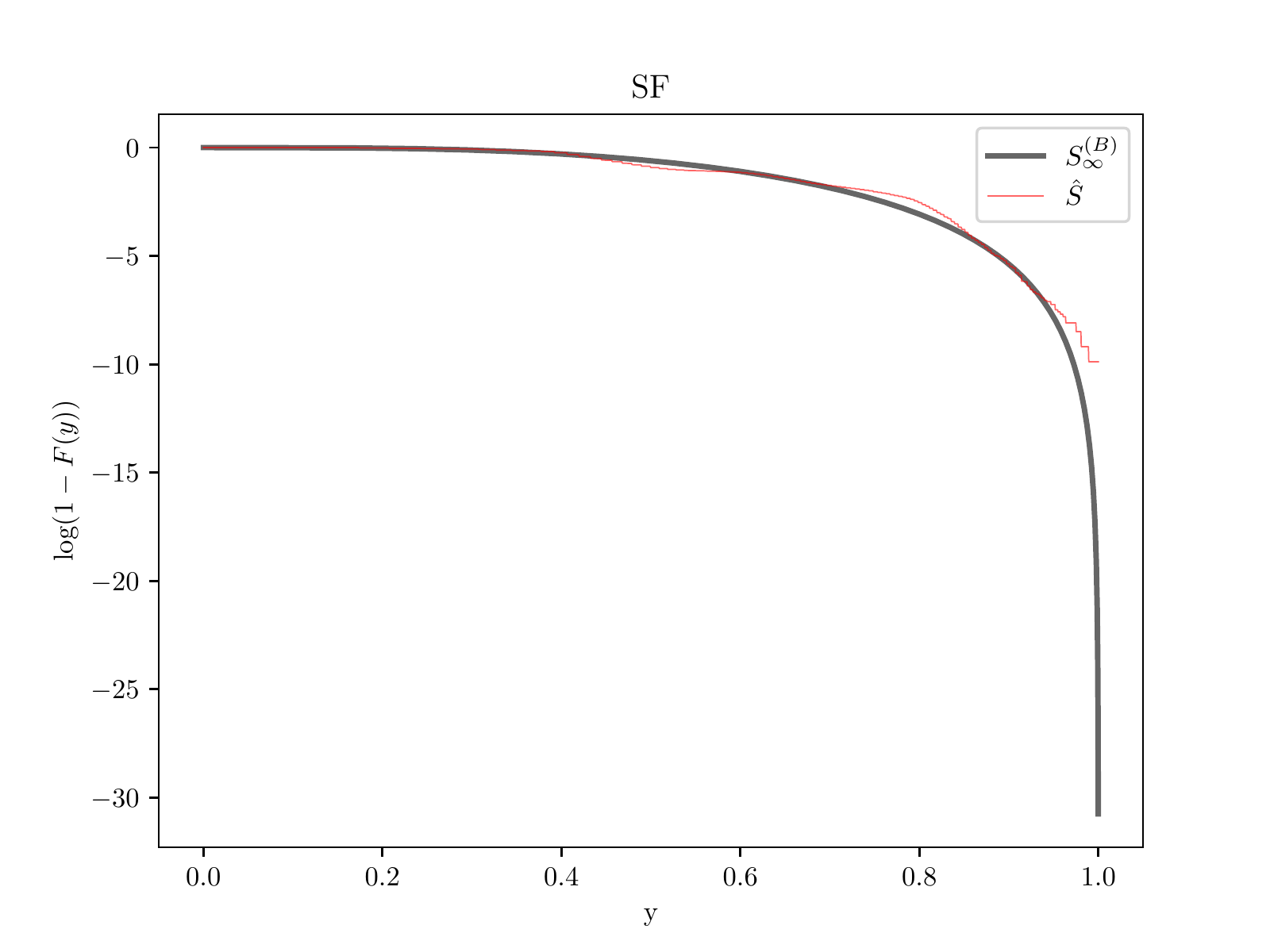}
    \caption{Plots of the estimated SF  $\hat{S}$ (red), and $S^{(B)}_\infty$ (black) as derived from the fitted parameters}
    \label{fig:SF}
\end{figure}
Figure \ref{fig:SF} reports the comparison within the two curves.
We note that the estimated SF $\hat{S}$ stops early w.r.t $S^{(B)}_\infty$.
This is due to the fact that the last evaluation performed in $\hat{S}$ before reaching 1 is in 0.999949.
It turns out that $\log(1-\hat{F}(0.999949)) = -9.8838$ that is the last computable value different from 1.

Note that, by fixing the value of $\delta$ ($\delta = 1/4$ in the present case), we can obtain from \fer{be} and\fer{new-be} the values of $p$ and $\lambda$ in terms of $\alpha$ and $\beta$. It holds
\be\label{values}
p = \left( 1 + \frac\beta{\alpha +3}\right)^4; \quad \lambda = \frac4{\alpha +\beta+3}.
\ee
Thus, by substituting the values in \fer{true} we obtain
\be\label{pp}
p = 5.477818, \quad \lambda= 0.373565.
\ee
Consequently, as $w_T p$ represents the maximum  weight  allowed, we can estimate from this the target value $w_T$.
 
To this extent, we evaluate the mean value of the population dataset. Taking into account the relationship between the generalized Beta distribution and the Beta distribution, in the case $\delta = 1/4$ we have
 \[
 \int_0^{pw_T} w\,f_\infty(w) \, dw = pw_T\frac{B(\alpha +4,\beta)}{B(\alpha,\beta)} = pw_T\frac{\alpha(\alpha +1)(\alpha+2)(\alpha+3)} {(\alpha+\beta)(\alpha +\beta +1)(\alpha+\beta +2)(\alpha+\beta +3)}.
  \]
 Thus, substituting in the expression above the values of $\alpha, \beta$ and $p$, as given by \fer{true} and \fer{pp}, we obtain
 \be\label{mean1}
  \int_0^{pw_T} w\,f_\infty(w) \, dw = 0.659217 \cdot w_T,
 \ee
which shows that the mean value of the weight is strictly lower than the target value $w_T$.
We can now  estimate the value of $w_T$ simply by evaluating the \emph{mean value} of the data in the original scale. 
Recalling that $\hat{\bm{w}}$ is the vector of the observed weights, we obtain
\[ \frac{1}{n}\sum_{i = 1}^n \hat{w_i} = 29.7099.\]
Hence, the target weight $w_T$ for the Chiroptera population takes the value
 \[
w_T = 45.068467. 
 \]
Also,  the limit weight allowed for the population is given by
\[
pw_T = 246.876863,
\]
which is almost twice the weight of the heaviest adult of Chiroptera in the dataset.
Last, considering that the variance  $\sigma$ can be obtained from \fer{hetero}, where the value of the constant $a= \mu/(2\delta)$ can be obtained by energetic considerations \cite{West}, assuming a value $a= 0.23$ similar to the growth rate of rats we obtain for $\mu$ and $\sigma$ the values
\be\label{sigma}
\mu = 0.115, \quad \sigma = 0,17.
\ee

%This value is less precise w.r.t the previous one.
%
%This can be due to the fact that our dataset contains different species of Chiroptera with a weight that lies in a wide range.
%
%However, the dimension of the sample of each specimens may not be representative of the true distribution of such mammalians in the planet. 
% 
%We also imagine that Chiroptera with a small weight are more present worldwide, but also easier to capture and measure w.r.t larger ones.
%
%This can motivate the fact that the sample mean is less w.r.t the theoretical mean.
%
%We imagine that, by having the a sample that contains the \emph{true fractions} of samples for each species, the two means should barely coincide.
%
%Unfortunately, this data is hard to obtain.
%

%\textcolor{red}{Al contrario di quello che pensavo, temo che questo sia un dato non reperibile. Se serve pittosto, come una sorta di appendice, posso farlo anche per un'altra specie non solo per il Chirotteri, per fare vedere diciamo che la strategia funziona ``in generale''}

%%%%%%%%%%%%%%%%%%%%%%%%%%%%%%%%%%%%%%%%%%%%%
 
\section{Conclusions}

The statistical distribution of body size in Chiroptera population has been described in this    paper by resorting to classical methods of collision-like kinetic theory, along the line of the interesting description of some evolution phenomena discussed by Galton in his famous book \emph{Natural Ihneritance} \cite{Gal2}. The main goal of our analysis consists of providing a possible explanation of the emergence of steady profiles for the body size distribution in the form of generalized Beta distributions of the first kind. The macroscopic behavior is consequent to the choice made at the microscopic level, choice that takes into account the essential features of the growth behavior in mammals \cite{West}.  The kinetic modeling is similar to the one introduced in Ref. \cite{GT1}, subsequently generalized in Refs. \cite{DT,GT2}, in which the human behavior is responsible of the formation of a macroscopic equilibrium in the form of probability distributions with thin tails, like the lognormal or the generalized Gamma distributions. From this point of view, the present results can be considered as an extension of the kinetic description of social phenomena considered in Refs. \cite{DT,GT2,Tos2}, which allows to quantify, at a macroscopic level, the effect of constraints in the elementary interaction which produce in the long time limit a whole class of distribution with bounded support in the form of generalized Beta distributions of first kind.  Well-known arguments of kinetic theory allow to model these phenomena by means of a Fokker--Planck equation with variable coefficients of diffusion and drift. In view of its steady state profile, convergence to equilibrium for this new Fokker--Planck equation could be studied by resorting to the methods recently developed in Ref. \cite{FPTT3} to quantify the rate of convergence  towards equilibrium for a similar equation describing opinion formation.   

%%%%%%%%%%%%%%%%%%%%%

\section*{Acknowledgement} 
This work has been written within the activities of GNFM and GNCS groups of INdAM (National Institute of High Mathematics). The research was partially supported by MIUR - Dipartimenti di Eccellenza Program (2018-2022) - Dept. of Mathematics  "F. Casorati" University of Pavia. 

%%%%%%%%%%%%%%%%%%%%%%%%%%%%%%%%%%%%%%%%%%%%%%%%%%%%%%%%%%%%%%%%%%%%%%%%%%%%%%%

\end{document}